\documentstyle[preprint,aps]{revtex}

\def\be{\begin{equation}}
\def\ee{\end{equation}}
\def\bea{\begin{eqnarray}}
\def\eea{\end{eqnarray}}

\def\gm{$\Gamma$}

\begin{document}
\tightenlines
\draft
\preprint{UOM/NPh/HQP/00-1(revised)}
\title{A note on spectator effects and quark-hadron duality in \\
inclusive beauty decays}

\author{S. Arunagiri}
\address{Department of Nuclear Physics, University of Madras,\\
Guindy Campus, Chennai 600 025, Tamil Nadu, INDIA}
\date{\today}
\maketitle

\begin{abstract}
In this paper, we evaluate the expectation values of four-quark operators
for the inclusive beauty decays from the differences in the inclusive
total decay rates, assuming that the heavy quark expansion 
converges at $O(1/m^3)$. The obtained expectation values 
yeilds the ratio $\tau(\Lambda_b)/\tau(B)$ close to 
the experimental one. We further point out that the quark-hadron
duality violation would be rather small allowing predictions of inclusive
quantities. 
\end{abstract}
\hskip0.5cm
\section{introduction}

According to heavy quark expansion (HQE) \cite{bigi},
inclusive decay rates of heavy hadrons are expected to
have almost same lifetimes.
At the leading order of the HQE, the hadronic decay rate is,
that of the free heavy quark decay, proportional to $m^5$, where
$m$ is the heavy quark mass. The decay
rate at the next-to-leading order (NLO) includes the terms of
heavy quark motion
inside the hadron and chromomagnetic mass splitting due to
spin orientation of the heavy quark. Since the latter vanishes for
baryons except for $\Omega_Q$, the NLO decay rate splits up into
mesonic and baryonic ones. However, the lifetimes of baryon
and meson of given flavour is expected to be roughly the same. But, 
the experimental value of the ratio 
$\tau(\Lambda_b)/\tau(B) = 0.78$, much lower than the theoretical prediction of
0.9. In order to reduce the discrepancy, the contributions from the terms
of $O(1/m^3)$ have necessarily to be included. At the third order in $1/m$,
the decay rate becomes flavour dependent, involving both the heavy
and light quark fields. The matrix element at order $1/m^3$ is
\be
C_6(\mu){1 \over {2M}}\left<H|(\bar Q \Gamma^{\mu} Q)(\bar q \Gamma_{\mu}
q)|H\right>
\ee
where the coefficient functions, $C_6$, describe
the spectator quarks processes such as Pauli interference (PI),
weak annihilation (WA) and $W$-scattering (WS).
The evaluation of the expectation values of the operators in the above equation 
is not straight. The difficult tast of this evaluation is nevertheless 
accomplished taking course to vacuum saturation approximation (in the case
of meson), valence quark approximation (in the case of baryon), hadronic 
parameterisation and QCD sum rules \cite{rosner,neubert,colangelo}. 
The result showed that the third order
term, precisely the fourquark operators, does not adequatly
contribute to enhance the decay rate of $\Lambda_b$.
In \cite{volo}, Voloshin pointed out the relations between the decay
rates of the heavy baryons. But these relations are not applicable
to the cham sector for the reason mentioned below. 

In this paper, we attempt to evaluate the expectation values of four-quark
operators from the differences in the total decay rates of the $b$-hadrons. 
That means we assume that the heavy quark expansion, being asymptotic in
nature, converges at $O(1/m^3)$ of the expansion. Already the next-to-leading
order contribution due to terms of $O(1/m^2)$ is only about five percent
of the leading order. Thus, it cannot be anticipated that the size
of the terms at the third order in $1/m$ more than a few percent.
On the other hand, the observation made above is applicable only to the 
beauty case,
since
\be
{16 \pi^2 \over {m_c^3}} C(\mu) \left< O_6 \right>_{H_c}   \gg
{16 \pi^2 \over {m_b^3}} C(\mu) \left< O_6 \right>_{H_b}   
\ee
where $C(\mu)$ stands for some structure involving $c_\pm$ and 
$\left< O_6 \right>_H$ the dimension six four-quark operators (FQO) of hadron.
Hence, in the background described, we make use of the total decay 
rates to obtain the expectation values of four quark operators (EVFQO) for
$b$-hadrons. Thus, the present evaluation depends only
on the heavy quark expansion and the $SU(3)$ flavour symmetry.

\section{splitting of decay rates and expectation values of 
four-quark operators}
The $B$ mesons, $B^-$, $B^0$ and $B^0_s$, are triplet
under $SU(3)_f$ flavour symmetry. Their total decay rate
splits up due to its light quark flavour dependence at the third order
in the HQE. The differences in the decay rates of the triplet,
\gm($B^0_d$) - \gm($B^-$), \gm($B^0_s$) - \gm($B^-$) and
\gm($B_s^0$) - \gm($B^0_d$), are related to the third order terms
in $1/m$ by
\bea
\Gamma(B^0_d)-\Gamma(B^-) &=& -\Gamma^\prime_0 (1-x)^2\nonumber\\
&& \times \left\{Z_1{1 \over 3}(2c_+ - c_-+6)+2(c_+-c_-/2+1)\right\}
\left<O_6\right>_{B^0_d-B^-} \label{one}\\
\Gamma(B^0_s) - \Gamma(B^-) &=& -\Gamma^\prime_0 (1-x)^2\nonumber\\
&& \times \left\{Z_2{1 \over 3}(2c_+ - c_-+6)+2(c_+ - c_-/2+1)\right\}
\left<O_6\right>_{B^0_s-B^-}\\
\Gamma(B^0_s)-\Gamma(B^0_d) &=& -\Gamma^\prime_0 (1-x)^2\nonumber\\
&& \times \left\{(Z_1-Z_2){1 \over 3}(2c_+-c_-+6)\right\}
\left<O_6\right>_{B^0_s-B^0_d} \label{three}
\eea
where $\Gamma^\prime_0$ = $2G_f^2|V_{cb}|^2m_b^2/3\pi$,
$c_- = c_+^{-2} = (\alpha(m)/\alpha(m_W))^{2\gamma/\beta}, 
\beta = 11-2n_f/3$ ($n_f$ being the number of flavour), $\gamma = 2$, 
$x$ = $m_c^2/m_b^2$ and
\bea
Z_1 &=& \left(cos^2\theta_c(1+{x \over 2})+
sin^2\theta_c \sqrt{1-4x}(1-x)\right)\\
Z_2 &=& \left(sin^2\theta_c(1+{x \over 2})+
cos^2\theta_c \sqrt{1-4x}(1-x)\right)\\
\left<O_6\right> &=& \left<{1 \over 2}(\bar b \Gamma_{\mu} b)
[(\bar d \Gamma_{\mu} d)
-(\bar u \Gamma_{\mu} u)]\right> \equiv 
\left<{1 \over 2}(\bar b \Gamma_{\mu} b)[(\bar s \Gamma_{\mu} s)
-(\bar q \Gamma_{\mu} q)]\right>
\eea
with $q = u, d$.
In eqns. (\ref{one}-\ref{three}), 
the $rhs$ contains the terms corresponding to the
unsuppresed and suppressed nonleptonic decay rates and
twice the semileptonic decay rates at the third order.

For the decay rates \gm($B^-$) = 0.617 $ps^{-1}$,     
\gm($B^0$) = 0.637 $ps^{-1}$ and \gm($B^0_s$) = 0.645 $ps^{-1}$ \cite{pdg},
the EVFQO are obtained for $B$ meson, as an average from
eqs. (\ref{one}-\ref{three}):
\be
\left<O_6\right>_B = 8.08 \times 10^{-3} GeV^3.
\label{bval}
\ee
This is smaller than the one obtained
in terms of the leptonic decay constant, $f_B$.

On the other hand, for the triplet baryons,
$\Lambda_b$, $\Xi^-$ and $\Xi^0$, with
$\tau(\Lambda_b)$ $<$ $\tau(\Xi^0) \approx \tau(\Xi^-)$,
we have the relation between
the difference in the total decay rates and the terms of
$O(1/m^3)$ in the HQE, as
\be
\Gamma(\Lambda_b)-\Gamma(\Xi^0) = {3 \over 8}\Gamma_0^\prime (-c_+(2c_-+c_+-2)
\left<O_6\right>_{\Lambda_b-\Xi^0}
\ee
We obtain the EVFQO for
the baryon
\be
\left<O_6\right>_{\Lambda_b-\Xi^0} = 3.072 \times 10^{-2} GeV^3
\label{lval}
\ee
where we have used the
decay rates corresponding to the lifetimes 1.24 $ps$ and 1.39 $ps$ \cite{pdg}
of $\Lambda_b$ and $\Xi^0$ respectively.
The EVFQO 
for baryon is about 3.8 times larger than that of B. But, it is about
3.5 - 4.0, dpending on the 
changes in the mass and other sources of uncertainity.For
these values
\be
\tau(\Lambda_b)/\tau(B) = 0.81 - 0.84 
\ee
Using the experimental value of $\tau(B^-)$ = 1.55 $ps$
alongwith the above theoretical value,
the lifetime of $\Lambda_b$ turns out to be
\be
\tau(\Lambda_b) = {\Gamma(\Lambda_b) \over {\Gamma(B)}} \tau(B^-) =
1.26 - 1.3~ps.
\ee
The central values would change little bit in view of the 
uncertainities in the parameters like the heavy quark mass and the 
kinetic energy paprameter.

The expectation values for baryon is quite large. In fact, Rosner \cite{rosner}
found it to be 1.8 times larger than that of meson which accounts
for about 25\%, 45\% if renormalisation group improved ,
of the needed enhancement in the decay rate. However, what we obtain
is model independent estimate. We have only used the experimental total decay
rates. The result is surprising as well as genuine, if one has to believe
the experimental values used. Still, there is room for uncertainity
of few percent. On the other hand, the generic structure we employed
can decomposed into as found in \cite{neubert}
by Neubert and Sachrajda. Using them will give an improved estimate
but in that case it may become somewhat model dependent. 

We will briefly take note two previous works. In \cite{arun}, 
using potential model,
the expectation values of four-quark operators of meson and baryons are
obtained and the ratio $\tau(\Lambda_b)/\tau(B)$ is estimated to be in the
range 0.79 - 0.84. Using QCD sum rules \cite{liu}, assuming a different duality
ansatz, in  the ratio is got to be 0.81. Though these methods arise on 
different basis, they seem to agree with the present model independent
evaluation.
 
\section{quark-hadron duality: a comment}
Assuming the convergence of the heavy quark expansion is valid one 
as long as there is an uncertainity of few percent. This would also 
positively suggest that the quark-hadron duality violating oscillating 
component might be small. So such violation would not deter a decent 
determination of quantities of interest in the heay quark expansion
like the lifetimes of beauty hadrons, the semileptonic branching ratio
and the charm counting.
For the reasons mentioned in the beginning, the assumption on convergence 
cannot be made for charmed case. 

There are renormalon contribution from the perturbative part of the expansion
\cite{zak}.
They are IR renormalons of the order $\Lambda_{QCD}$. We, as of now, don't
have deep insight of it. They would be expected to differ for meson and
baryon. Because this contribution does not correspond to any local
operators of the thoery and independent of the heavy quark mass, 
a difference of about 50 to 100 MeV between 
meson and baryon would imply much significance for the quatities
described by the heavy quark expansion. We cannot on obviuos terms
argue that renormalons are related to the assumption of quark-hadron duality.
On the other hand, it will shed light on the quantities concerned and the
underlying assumption versus the heavy quark expansion. 

\section{concluding remarks}
We have estimated the expectation values of the four-quark
operators of beauty hadron and found that they yielded the ratio of lifetimes
of $\Lambda_b$ and $B$ meson closer to the experimental value. We conclude 
by noting that the model indepent prediction will be an accurate one when 
other structure of the currents given in \cite{neubert}
are also included. 

\acknowledgements
The author is grateful to Prof. H. Yamamoto,
Prof. P. R. Subramanian, Dr. D. Caleb Chanthi Raj, 
and Mr R. Justin Joseyphus
for discussions and encouragement. He acknowledges
UGC for the support through its Special Assistance Programme
(Phase III).

\end{document}